\documentclass{jaa}
\usepackage{graphicx}
\usepackage{xcolor}
\newcommand\arcdeg{\mbox{$^\circ$}}%
\newcommand\arcmin{\mbox{$^\prime$}}%
\newcommand\arcsec{\mbox{$^{\prime\prime}$}}%
\newcommand\farcdeg{\mbox{$.\!\!^\circ$}}%
\newcommand\farcmin{\mbox{$.\mkern-4mu^\prime$}}%
\newcommand\farcsec{\mbox{$.\!\!^{\prime\prime}$}}%
\newcommand\micron{\mbox{$\mu$m}}%
\newcommand\etal{{\em et al. }}%
\begin{document}\sloppy

%%paper title
%%For line breaks \\ can be used within title 
\title{ Observations with the 3.6 meter Devasthal Optical Telescope}

%%author names are separated by comma (,) 
%%use \and before the last author name 
%%use a * along with the number separated by comma
%% for the  author for correspondence
%%\textsuperscript{number} is used for affiliation
%%\affilOne, \affilTwo etc., upto \affilTwentyfive is possible
%%Please note the first letter after \affil is capitalised in the command
%%

\author{Ram Sagar\textsuperscript{1,2}, Brijesh Kumar\textsuperscript{2} and Saurabh Sharma\textsuperscript{2}}
\affilOne{\textsuperscript{1}Indian Institute of Astrophysics, Sarajapur road, Koramangala, Bengaluru 560 034, India.\\}
\affilTwo{\textsuperscript{2} Aryabhatta Research Institute of Observational Sciences, Manora Peak, Nainital 263 001, India.}

\twocolumn[{

\maketitle

%%include \corres to print the corresponding author Email id
\corres{ramsagar@iiap.res.in}

%%include \msinfo for
%%manuscript information such as
%%received, revised and accepted dates
%%
\msinfo{20 August 2020}{....... 2020}

%%abstract
\begin{abstract}
 The 3.6 meter Indo-Belgian Devasthal optical telescope (DOT) has been used for optical and near-infrared (NIR) observations of celestial objects. The telescope has detected stars of $B = 24.5\pm0.2, R = 24.6\pm0.12$ and $ g =25.2\pm0.2$ mag in exposure times of 1200, 4320 and 3600 seconds respectively. In one hour of exposure time, a distant galaxy of 24.3$\pm$0.2 mag and point sources of $\sim$ 25 mag have been detected in the SDSS $i$ band. The NIR observations show that stars up to $J = 20\pm0.1, H = 18.8\pm0.1$ and $K = 18.2\pm0.1$ mag can be detected in effective exposure times of 500, 550 and 1000 sec respectively. The $nbL$ band sources brighter than $\sim$ 9.2 mag and strong ($\geq$ 0.4 Jy) $PAH$ emitting sources like Sh 2-61 can also be observed with the 3.6 meter DOT. A binary star having angular separation of 0\farcsec4 has been resolved by the telescope. Sky images with sub-arcsec angular resolutions are observed with the telescope at wavelengths ranging from optical to NIR for a good fraction of observing time. The on-site performance of the telescope is found to be at par with the performance of other similar telescopes located elsewhere in the world. Due to advantage of its geographical location, the 3.6 meter DOT can provide optical and NIR observations for a number of front line Galactic and extra-galactic astrophysical research problems including optical follow up of GMRT and AstroSat sources and optical transient objects. 
 
\end{abstract}
%%insert keywords separated by 3 hyphens using \keywords{words}
 }]
%%close the twocolumn escape here

%%include \doinum{number}for the DOI number in the header
%%include \volnum{number} for the volume number in the header
%%include \year{yyyy} for  year of publication in the header
%%include \pgrange{num--num} page range of article in the header
%%include \artcitid{num} for the article citation id
%%include \lp to print last page of the article
%%include \setcounter{page}{pagenum} for the exact starting page of the article

\doinum{12.3456/s78910-011-012-3}
\artcitid{\#\#\#\#}
\volnum{000}
\year{0000}
\pgrange{1--}
\setcounter{page}{1}
\lp{1}

\section{Introduction}

\begin{figure*}
%\caption{caption spanning two columns}
%\centering
\includegraphics[width=.8\textheight, height=1.0\textheight]{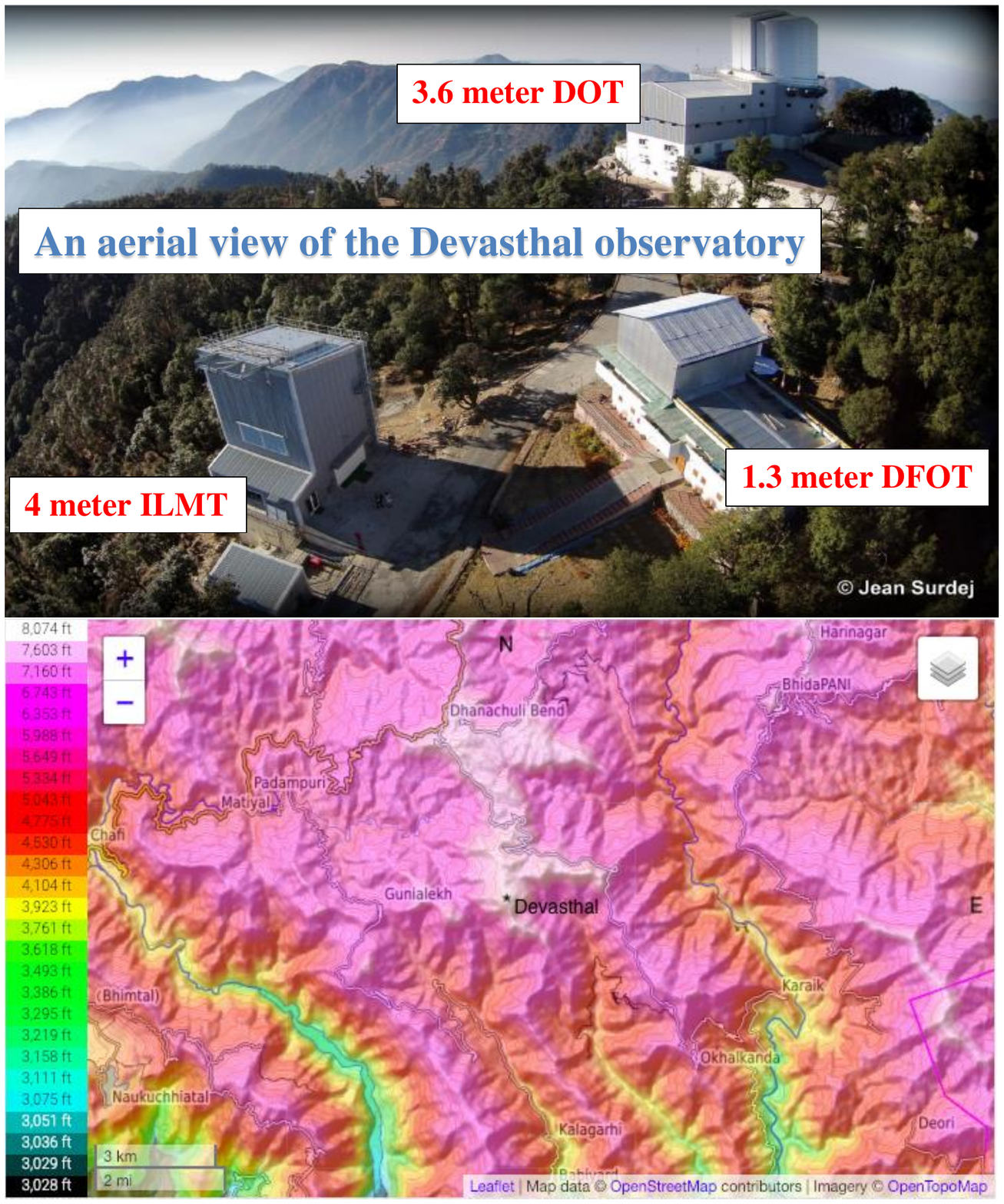}
\vspace{-5cm}
\caption{ Bottom part shows topographic contour map of the Devasthal and its immediate surroundings. North is up and East is to the right in this map. Devasthal, highest peak in the region of $\geq$ 10 km range, is at a point which has sharp altitude gradient towards the south-west, the prevailing incoming wind direction at the site. This location is therefore expected to provide laminar air flow resulting better seeing for astronomical observations. An aerial view of the Devasthal Observatory is shown in the upper part. Buildings of 3.6 meter DOT, 1.3 meter Devasthal fast optical telescope (DFOT) (Sagar \etal 2011) and 4 meter International Liquid Mirror Telescope (ILMT) (Surdej \etal 2018) are marked.}
\label{fig:site}
\end{figure*}

The Devasthal (meaning ‘abode of God’) is a mountain peak (longitude = 79\farcdeg{7} E, latitude = 29\farcdeg{4} N, and altitude = 2424$\pm$4 m). It is located at a distance of $\sim$ 55 Km by road from Nainital in Kumaon region of central Himalaya.  Figure~\ref{fig:site} shows an aerial view of the Devasthal observatory and topographic contour map of the region. The location was identified after decades of detailed site survey using modern instruments (Sagar \etal 2000; Stalin \etal 2001 and references therein). The 3.6 meter Indo-Belgian Devasthal Optical telescope (DOT) was successfully installed and technically activated by both the premiers of India and Belgium from Brussels on March 30, 2016 and since then it is in use for training people, testing back-end instruments and taking observations of various types of celestial objects.  

A description of the actively supported modern optical telescope is given in the next section. Back-end instruments presently in use and results of sky performance derived from the optical and near-infrared (NIR) observations taken with the telescope are given in the remaining sections. Last section provides summary and future out look of this international observing facility.

%\begin{figure*}
%\caption{caption spanning two columns}
%\centering\includegraphics[height=.25\textheight]{telescope-building-distance-view.jpg}
%\caption{ A long distance view of the 3.6 meter DOT telescope building along with buildings of 130-cm Devasthal fast optical %telescope (Sagar \etal 2011) and 4-m International Liquid Mirror Telescope (Surdej \etal 2018). It is located at an aerial distance %of $\sim$ 22 Km from Manora peak, Nainital, the main campus of ARIES.}
%\label{fig:telbuild}
%\end{figure*}

\section{ The 3.6 meter DOT}

The modern actively supported 3.6 meter DOT is a f/9 two mirror Ritchey-Chretien system. It has three Cassegrain ports with a back focal distance of 2.5 m. Details of history, technical, construction of building, mirror coating and installation of the telescope etc. are given by Omar \etal (2017); Kumar \etal (2018), Omar \etal (2019c) and Sagar \etal (2019a) along with a few initial scientific results. In order to detect and correct deformations, aberrations or any other phenomenon that degrade the image quality of the telescope, the 3.6 meter DOT is equipped with an active optics system (AOS) which compensates for distorting forces that change relatively slowly, roughly on timescales of seconds. The AOS consists of a wave front sensor (WFS), primary (M1) mirror support system consisting of 69 actuators generating forces on M1 mirror and 3 axial definers with load cells on M1 mirror, secondary (M2) mirror hexapod that supports M2 mirror and the telescope control system (TCS) which acts as interface between each elements of the telescope. Manufacturing surface inaccuracies of mirrors and imperfection in integration of the mirrors in their cells, gravity load, thermal effects and wind effects are the main sources of telescope image degradation. A WFS is used to measure and analyze the wave front coming from the telescope system. Its output is used to improve telescope image. The astigmatism, 3-fold and spherical aberration are corrected with the actuators of M1 mirror support while focus, coma and tilt can be corrected with the M2 mirror hexapod. Load cells measure the residual forces on the 3 axial definers and the actuators are used to keep these forces zero. The repeatable corrections on M1 and M2 mirrors are applied in open loop mode (look up table), whereas the close loop mode applies both repeatable and non-repeatable corrections. The later arises from thermal deformations and wind effects. The telescope uses sophisticated and complex techniques for achieving and maintaining the image quality while tracking the objects in sky. The telescope has, thus, online optics alignment while taking images of celestial objects and differs from classical telescopes where AOS is not used. 

The TCS accepts coordinates of both target and the guide star. Acting as the interface between the hardware of telescope and the user, the TCS provides access to both the operational and engineering control of the telescope hardware. It also interfaces with the AOS, guiding unit system and the back-end instrument. Further details on this are given by Kumar et al. (2018) and  Sagar et al. (2019a).

\subsection{ Imaging of close binaries}

\begin{figure*}
\centering\includegraphics[height=.25\textheight]{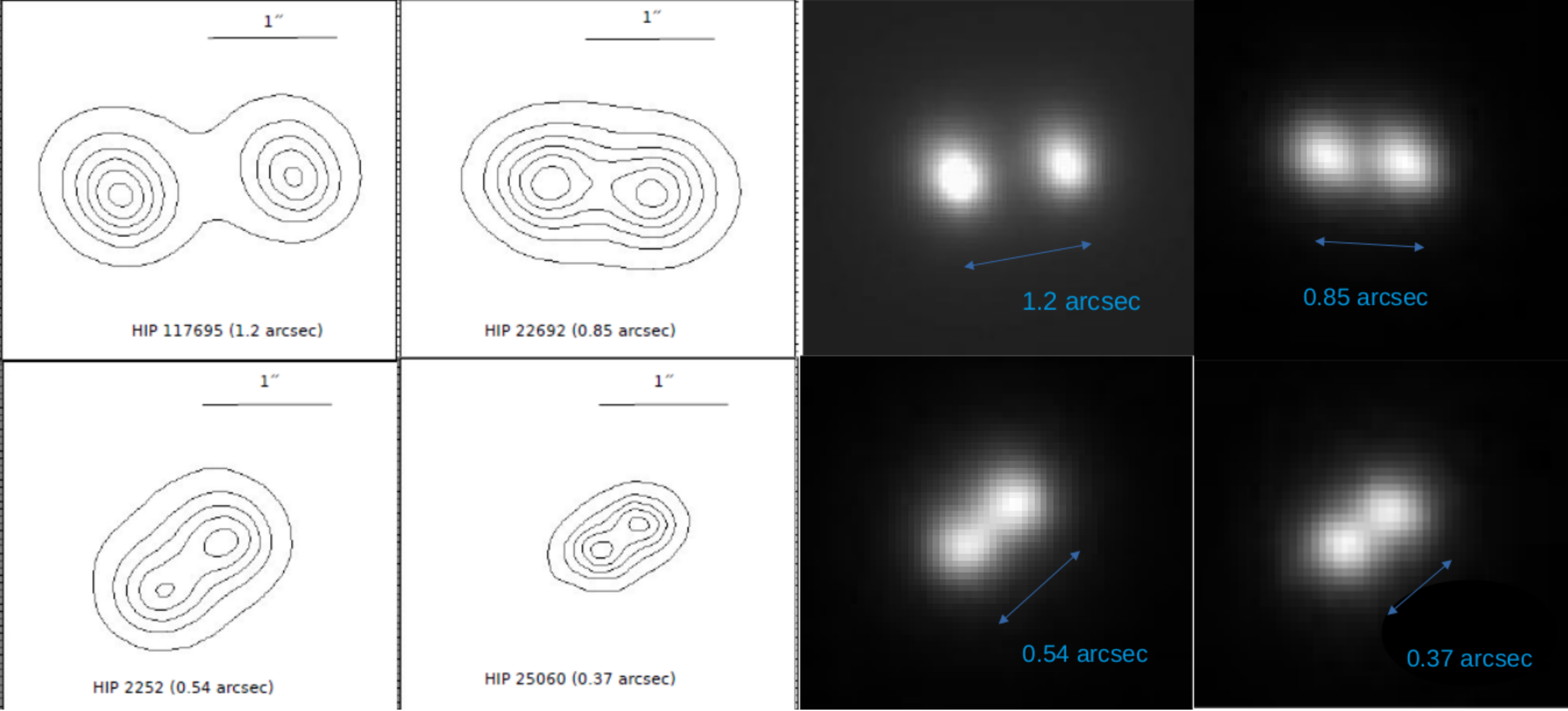}
\caption{ The Iso-intensity contour and images of 4 binary stars are shown. The images of binary stars having known angular separation between 0\farcsec37 to 1\farcsec2 are taken with the test-camera mounted at the axial Cassegrain port of 3.6 meter DOT. This shows that binary star having angular separation of 0\farcsec37 is well resolved. }
\label{fig:star}
\end{figure*}

During commissioning, a test-camera viz. an air-cooled Microline ML 402ME CCD chip of 768$\times$512 pixels was used to quantify performance of the 3.6 meter DOT. It has a pixel size of 9 $\micron$ which corresponds to 0\farcsec{06} at the focal plane of the telescope. During November-December 2015, the test-camera imaged few binary stars of known angular separation in broadband (0.45 to 0.6 $\micron$) visual glass filter. These images revealed that binary stars having separation of sub-arc-sec are clearly resolved (Figure~\ref{fig:star}). Best angular resolution of $\sim$ 0\farcsec4 was achieved in the night of 30th November 2015. More information on these observations are given by Kumar \etal (2018),

\subsection{Allotment of observing time }

Observing proposals from the users are submitted online twice in a year for both observing cycles namely Cycle-1 (February to May) and Cycle-2 (October to January). The website\footnote{\it https://www.aries.res.in} provides relevant information regarding the policies and procedures followed in allotment of observing time for the telescope. Based on scientific merit of the submitted proposals, the Belgian and Indian time allocation committees allot observing time to the proposers of their countries. Presently, 33 and 7 \% observing time are allotted to the proposers from ARIES and Belgium respectively while remaining 60 \% observing time is allotted to other proposers. An updated information on existing back end instruments on the 3.6 meter DOT are given below.

\section{Cassegrain port instruments} 

\begin{figure*}
\centering\includegraphics[height=.35\textheight]{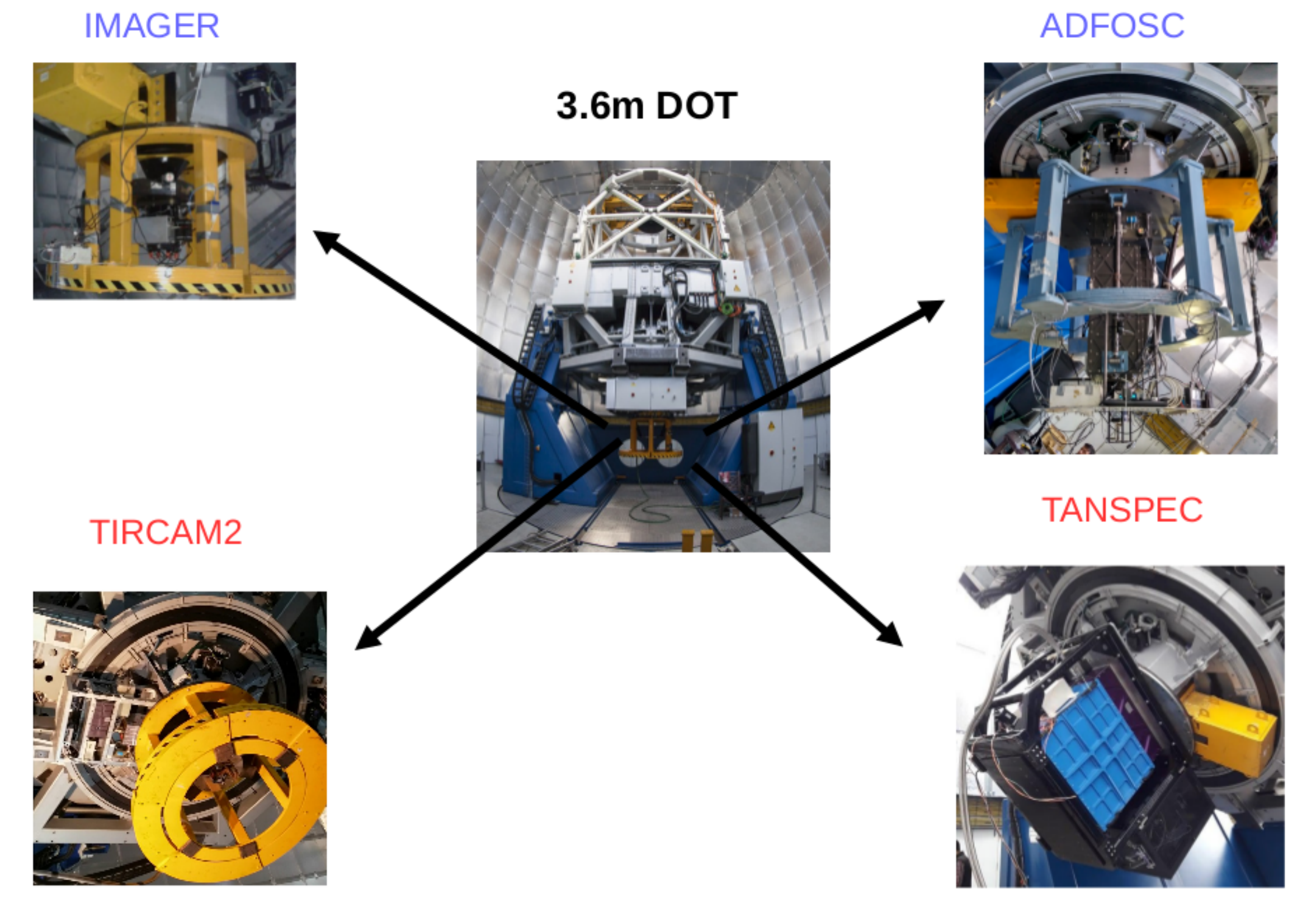}
\caption{ The 3.6 meter DOT is in the center. The pictures of back-end instruments 4K$\times$4K CCD IMAGER, ADFOSC, TIRCAM2 and TANSPEC used for regular observations are shown. }
\label{fig:ins}
\end{figure*}

The telescope has capacity to bear imbalance of 2000 Nm on altitude axes and of 400 Nm on azimuth axes. Imbalances are adjustable using motorized weights on altitude axes and fixed weights on azimuth axes. There are three Cassegrain ports available for mounting back-end instruments on the 3.6 meter DOT. The main axial port is designed for mounting instruments weighing up-to 2000 kg. The telescope interface plate (TIP) is used for mounting axial port instrument. Sagar \etal (2019a) have described in detail shape and  dimensions of the TIP. The center of gravity (CG) of axial port instrument is 80 cm below the TIP. One cm shift of CG position in the vertical axis creates a torque of 200 Nm.This value is very close to the altitude bearing friction. Hence, mounting of any axial port instrument needs careful adjustment of its CG.The side port instruments can have a weight of 250 kg each with CG of 62 cm away from the centre of TIP. The main axial and two side instrument ports are fixed to a structure, called ARISS. The device which rotates the image of sky and side-port fold mirror are also part of the ARISS. All these units are nested at the rear of M1 mirror. 
Figure~\ref{fig:ins} shows the pictures of the telescope along with back-end instruments presently in use for regular observations. The IMAGER and ADFOSC (ARIES-Devasthal Faint Object Spectrograph and Camera) are used in optical region while TIRCAM2 (TIFR NIR Imaging Camera-II) and TANSPEC (TIFR-ARIES NIR Spectrometer) are used in NIR region. The 3.6 meter DOT has been used for detailed characterisation of these instruments as well as for observations of proposals related to photometric and spectroscopic study of galactic star forming regions, star clusters, AGNs and quasars, supernovae, optical transient events and distant galaxies etc. The following sub-sections provide technical details of these back-end instruments.
 
 \subsection{ IMAGER : Optical imaging camera}

The IMAGER having  wavelength sensitivity between 0.35 \micron\ to 0.9 \micron, is indigenously designed, developed and assembled at ARIES. Pandey \etal (2018) have provided technical details of optical and mechanical design, motorized filter wheels and data acquisition system of the instrument. The pixel size of a blue-enhanced liquid Nitrogen cooled ($\sim -120$\arcdeg\ C) STA4150 4K$\times$4K CCD sensor is 15 \micron\ square. Standard broadband Bessel $U$, $B$, $V$, $R$, and $I$ and SDSS $u$, $g$, $r$, $i$, and $z$ filters are presently mounted. The FOV of IMAGER is $6\farcmin5 \times 6\farcmin5$ at the Cassegrain focus of the telescope. The IMAGER has been used extensively for taking images of both point and extended sources located in the sky. 

%\begin{figure*}
%\centering\includegraphics[height=.35\textheight]{fig8-new.pdf}
%\caption{ A sketch map of the instrument envelope for Cassegrain main axial and both side ports is shown. The actual dimensions of %the instrument envelope of the main axial port are marked. This figure is borrowed from Sagar \etal (2019a). }
%\label{fig:focal}
%\end{figure*}

\subsection{ ADFOSC : Optical imaging and spectroscopic camera}

The ADFOSC is a low-resolution slit spectrograph-cum-imager having wavelength sensitivity between 0.35 \micron\ to 0.9 \micron. It is also designed, developed and assembled in-house at ARIES. Omar \etal (2019a, b, c) have provided its detailed technical parameters along with its performance. Briefly, a collimator and a focal reducer are used to convert the telescope f/9 beam to a ADFOSC f/4.3 beam. The detector used is a closed-cycle cryogenically cooled grade-0 back-illuminated E2V 231-84 chip of 4096$\times$4096 square pixel CCD camera. The ADFOSC can be used in 3 modes of observations named as (a) broad and narrow band photometric imaging, (b)  long-slit low-resolution $({\lambda}/{\Delta}{\lambda} \sim 1000)$ and slit-less spectroscopy and (c) fast imaging (up to millisecond cadence) using an electron-multiplier frame-transfer CCD with smaller FOV and on-chip binning. In photometric imaging mode, it is equipped with SDSS $u$, $g$, $r$, $i$, and $z$ filters, 8\arcmin\ long slits, grisms, and narrow-band filters and can image a FOV of 13\farcmin6$\times$13\farcmin6 when mounted at the telescope. It has been used in both imaging and spectroscopy modes for the observations of stars, ionized star-forming regions and galaxies etc. (Omar \etal 2019a, b). 

 \subsection{TIRCAM2 : NIR imaging camera}

TIRCAM2, developed by the Tata Institute of Fundamental Research (TIFR) (Naik et al. 2012), is a closed cycle Helium cryo-cooled imaging camera equipped with a Raytheon $512\times512$ pixels InSb Aladdin III Quadrant focal plane array. This imaging camera has sensitivity from $\lambda =$ 1 \micron\ to 3.7 \micron. Pixel scale of the camera on the telescope is 0\farcsec17 with a FOV of 86\farcsec5$\times$86\farcsec5. The TIRCAM2 was mounted on the axial port of the telescope for tests, characterisation and science observations. Further details of this camera are given by Ojha \etal (2018) and Baug \etal (2018). It is equipped with standard $J$ (1.2 \micron), $H$ (1.65 \micron) and $K$ (2.19 \micron) broad $(\Delta \lambda \sim$ 0.3-0.4 \micron) photometric bands, and narrow $(\Delta \lambda \sim$ 0.03-0.07 \micron) band $B_{r-\gamma}$ (2.16 \micron); $K_{cont}$ (2.17 \micron); Polycyclic aromatic hydrocarbon $(PAH)$ (3.29 \micron) and narrow band L $(nbL)$ (3.59 \micron) filters. TIRCAM2 provides sampling time of $\sim$ 256 ms for the full frame and $\sim$ 16 ms for a sub-array window of $32\times32$ square pixels. TIRCAM2 is being used both for deep NIR imaging of celestial sources as well as for fast imaging in case of lunar/planet occultation events. Because of its observational capability up to 3.59 \micron, this camera is extremely valuable instrument for observing those bright nbL and PAH sources which are saturated in the Spitzer-Infrared Array Camera observations.

\subsection{ TANSPEC : Optical-NIR imaging and spectroscopic camera}

The TANSPEC, jointly developed by TIFR and ARIES, is an Optical-NIR medium resolution spectrograph. It covers $\lambda$ from 0.55 $\micron$ in optical up to 2.54 $\micron$ in NIR with a resolving power of $\sim$ 2750. It can be used for simultaneous observations across entire wavelength region. Optical lay out and technical details of this instrument are provided by Ojha \etal (2018). Briefly, it converts the f/9 telescope beam into f/12 beam on to a slit having range of widths from 0\farcsec{5} to 4\farcsec{0}. 
One pixel of the spectrograph 2048 $\times$ 2048 Hawaii-2RG (H2RG) array corresponds to 0\farcsec{25} and it operates in two modes. In highest resolution ($\sim$ 2750) mode, combination of a grating and two prisms are used while low resolution ($\sim$ 100-350) prism mode is used for high throughput observations. The instrument also has an independent imaging camera with a 1K$\times$1K H1RG detector which is the slit viewer. FOV of the slit viewer is 1\arcmin $\times$ 1\arcmin while its one pixel corresponds to 0\farcsec{25} on the sky. This camera  is used for the telescope guiding as well as for sky imaging. It also functions as a pupil viewer for instrument alignment on the telescope. It is equipped with both broad (r', i', Y, J, H, Ks) and narrow (H$_2$ \& B$_r$) band filters. The TANSPEC, after successful completion of laboratory test at Maunea-Kea Infrared, USA  was transported to the Devasthal. It was mounted and successfully tested as a back-end instrument on the 3.6 meter DOT during April-May, 2019. The initial results of performance tests of the TANSPEC are found to be very encouraging since they are at par with the design specifications. A detailed paper on these commissioning tests of the TANSPEC is under preparation.

\section{Factors affecting detection limit of a telescope}

The light gathering power of an optical telescope is related to its diameter ($D$) that collects and focuses the light. With the larger value of $D$, more photons are collected due to large area which makes it possible to study relatively fainter stars. For sky background limited observations (Sagar 2017), efficiency of a telescope to detect a celestial object at a frequency ($\nu$) is $\propto \sqrt{\frac{A_{\rm eff} \times I(t)} {\epsilon_{\rm D} \times B(\nu)}}$, where $A_{\rm eff}$ is the light gathering power of the telescope of diameter $D$ including the losses due to optics and the quantum efficiency of the detector used at the focus of the telescope; $B(\nu)$ is the sky background intensity at frequency $\nu$; I(t) is the integration time and $\epsilon_{\rm D}$ is the solid angle formed by the combination of atmospheric seeing and image degradation introduced due to optical and mechanical elements of the telescope including improper focusing as discussed earlier in Section 2. The light gathering power, $A_{\rm eff}$, of 3.6 meter DOT telescope mainly depends on reflectivity of both M1 and M2 mirrors, the losses due to optical components and the quantum efﬁciency of the detectors used in the back end instruments. The value of Sky background, $B(\nu)$, depends on both light pollution and lunar phase during observations. Detection limits of the 3.6 meter DOT at optical and NIR wavelengths are therefore estimated from observations taken on different epochs as described below.     

\begin{table*}[htb]
\tabularfont
\caption {The back-end instruments used for optical and NIR observations of the objects are given in the first and second column. Epoch, exposure time and filter used for observations are given in the 3$^{rd}$, 4$^{th}$ and $5^{th}$ columns respectively. The value of full width half maximum (FWHM) listed in $6^{th}$ column is estimated from stellar images. Sky brightness and magnitude of the object derived from the observations are listed in the last two column along with their associated error. } 
%\label{Table1}

\begin{tabular}{llllccll}  \topline
\textbf{Instrument}&\textbf{Object(s)}&\textbf{Epoch}&\textbf{Exposure} & \textbf{Filter(s)}&\textbf{FWHM} &\textbf{Sky brightness} & Magnitude\\
  &  & & (Seconds) & & ($\arcsec{}$) & mag/arcsec$^2$ &(mag) \\ \hline 
%\midline
 IMAGER & NGC 4147 & 23 March 2017& 1200& $B$& 1.2 &22.29$\pm$0.34 & 24.5$\pm$0.2 \\ 
 IMAGER & GRB 130603B& 24 March 2017&600& $B$& & & 22.13$\pm$0.05 \\  
 IMAGER & NGC 4147& 23 March 2017& 1200& $R$& 1.12 &19.36$\pm$0.21 & 23.5$\pm$0.2 \\ 
IMAGER & GRB 130603B& 24 March 2017&600& $R$& & & 20.72$\pm$0.02 \\
IMAGER &SN 2016B& 2 April 2017 &  & $R$& & & 19.79$\pm$0.05 \\
IMAGER & GRB 200412B & 23 April 2020 &4320 &$R$& $\sim$0.8 & &24.6$\pm$0.12 \\ 
IMAGER & GRB 200412B & 24 April 2020 &3600 &$g$ & $\sim$0.9 & &25.2$\pm$0.1 \\
ADFOSC & TGSS J1054+5832 &16 April 2018&3600 &$r$ & 1.0 &20.96$\pm$0.2&24.5$\pm$0.2 \\
ADFOSC & TGSS J1054+5832 &16 April 2018&3600 &$i$ & 1.0 &20.43$\pm$0.2&24.9$\pm$0.2 \\
ADFOSC & GRB 200524A & 24 May 2020 &300 &$r$ &  & &21.1$\pm$0.03 \\ 
TIRCAM2 & M 92 & 11-22 May 2017 & 550 & J &$\sim$0.8 & 16.4$\pm$0.2 & 19.0$\pm$0.1 \\
TIRCAM2 & M 92 & 11-22 May 2017 & 550 & H &$\sim$0.8 & 14.0$\pm$0.2 & 18.8$\pm$0.1 \\
TIRCAM2 & M 92 & 11-22 May 2017 & 1000 & K &$\sim$0.8 & 12.2$\pm$0.2 & 18.0$\pm$0.1 \\
TIRCAM2 & WISE Sources & 11-22 May 2017 & 25 & nbL & $\sim$0.8 &3.0$\pm$0.2 & 9.2$\pm$0.3 \\
TIRCAM2 & Czernik 3 &7 Oct 2017&500&J&$\sim$0.6&16.3$\pm$0.2 & 19.9$\pm$0.1 \\
TIRCAM2 & Czernik 3 &7 Oct 2017&500&H&$\sim$0.6&13.9$\pm$0.2 & 18.6$\pm$0.1 \\
TIRCAM2 & Czernik 3 &7 Oct 2017&1000&K&$\sim$0.6&12.2$\pm$0.2 & 18.2$\pm$0.1 \\
TIRCAM2 & NGC 7027 &13 Oct 2017&  &K&$\sim$0.8&  &  \\
TIRCAM2 & BD +36 \arcdeg 3639 &10 May 2018&  &K&$\sim$0.6&  &  \\
TIRCAM2 & IRC-20156 & 18 May 2018&  &K&$\sim$0.7&  &  \\
TIRCAM2 & SAO 98770 & 21 May 2018&  &K&$\sim$0.7&  &  \\
\hline
\end{tabular}
\end{table*}

\section{ Observations at optical wavelengths}

 Since March 2017, both point and extended types of celestial objects were imaged at optical wavelengths with the IMAGER and ADFOSC mounted at the main axial port of the 3.6 meter DOT (see Table 1). A brief description of these observations is given below.

 The $B$, $V$ and $R$ broadband images of the Galactic globular cluster NGC 4147 were obtained on 6 nights during 23 March 2017 to April 9, 2017. The sky brightness estimated from the images taken on 23 March 2017 (dark night) are 22.29$\pm$0.34 and 19.36$\pm$0.21 mag/arc-sec$^2$ in $B$ and $R$ bands respectively. These observations detect stars of $B =$24.5 mag and $R=$23.5 mag with S/N ratio of 5 in the deep colour magnitude diagram presented by Pandey \etal (2018). The number of short exposure images ranging from 30 to 50 seconds taken in $V$ and $R$ bands were 339 and 302 respectively (see Lata \etal 2019). The FWHM values of these images ranged from 0\farcsec{7} to 1\farcsec{0}. Based on these observations, Lata \etal (2019) identified and studied properties of 42 (including 28 newly discovered) periodic stellar variables.

 A few optical transient sources were observed with the IMAGER. A Short-duration GRB 130603B afterglow and its host galaxy were imaged 1387 days after the GRB event. The flux estimated from these observations have been used to construct the multi-wavelength spectral energy distribution of the host galaxy which indicates that the host galaxy is young and blue with moderate values of star-formation activities (Pandey \etal 2019). Kumar \etal (2020) reported broad band optical photometric observations of the GRB 200412B afterglow taken with the IMAGER during 15 to 25 April 2020. Multiple CCD frames, each having exposure times of 360 seconds, were taken in $I$, $R$ and $g$ bands. The GRB 200412B afterglow decayed $\sim$ 3 mag during the period of these observations. The magnitudes reported by Kumar \etal (2020) are listed in Table 1 along with other relevant information. Detailed analysis of these data is in progress. The type II supernova (SN) ASASSN-16ab/SN 2016B located in the galaxy PGC 037392 was observed 465 days after its explosion, in 2$\times$2 binning mode of the IMAGER (Dastidar \etal 2019). 
 
 Using ADFOSC, optical observations of a GMRT high red-shift (z $\sim 4.8\pm2$) radio galaxy source (TGSS J1054+5832) were obtained on a dark night in SDSS $r$ and $i$ pass bands. A statistically significant optical detection has been made in the $i$ band (Omar \etal 2019b). The magnitudes and associated errors of this object as well as sky brightness derived from these observations are listed in Table 1. In the $i$ band imaging of Abell cluster with ADFOSC, a source of 25$\pm$0.3 mag has been detected by Omar \etal (2019a, c) in an exposure time of $\sim$ 1 hour. Sanwal \etal (2020) observed GRB 200524A optical afterglow on 24 May 2020 with the ADFOSC. Several images of 300 seconds exposure time were taken in $g$, $r$ and $i$ bands. They clearly detect the optical transient and its $r$ band magnitude is listed in Table 1. Further processing of the data is in progress.
 
The night sky brightness values at Devasthal are found to be 22.29$\pm$0.34 and 19.36$\pm$0.21 mag/arc-sec$^2$ in the $B$ and $R$ photometric pass-bands respectively while they are $\sim$ 21 and $\sim$ 20.4 mag/arc-sec$^{2}$ in the SDSS $r$ and $i$ bands respectively (Table 1). These numbers indicate that the night sky at Devasthal is dark in optical region. The magnitude and associated error derived in different photometric pass bands for various objects are listed in Table 1. These numbers agree fairly well with the corresponding simulated values given by Pandey \etal (2018) for the IMAGER assuming values of 3600 seconds and 1\farcsec{1} for integration time and seeing respectively. Stars of $B = 24.5\pm0.2, R = 24.6\pm0.12$ and $ g =25.2\pm0.2$ mag have been detected in exposure times of 1200, 4320 and 3600 seconds respectively. In photometric imaging mode of the ADFOSC, a distant galaxy of 24.3$\pm$0.2 mag and point sources of $25\pm0.3$  mag have been detected in the SDSS $i$ band in one hour of exposure time. These detected magnitudes in different filters are $\sim$ 4 mag fainter than the corresponding magnitudes determined from similar observations taken with the 104 cm Sampurnanand telescope of ARIES located at Manora peak, Nainital (Sagar 2018). However, out of this 4 mag gain, only 2.78 mag can be attributed to the difference in aperture sizes of the telescopes. The remaining gain in the detection limit of the 3.6 meter DOT can be attributed to the dark sky and better seeing at Devasthal even after completion of the telescope and surrounding buildings.

 \section{ Observations at NIR wavelengths}
 
 The magnitudes estimated from TIRCAM2 observations taken during 11 to 22 May 2017 and on 7 October 2017 are used to quantify capability of the 3.6 meter DOT at NIR wavelengths. The magnitudes derived from these observations are listed in Table 1 along with other relevant information. They show that stars up to $\sim$ 20, 18.8 and 18.2 mag can be detected with 10\% photometric accuracy in the $J$, $H$ and $K$ bands respectively. The corresponding effective exposure times are 500, 550 and 1000 sec respectively. It is also capable of detecting the $nbL$ band sources brighter than $\sim$ 9.2 mag and strong ($\geq$ 0.4 Jy) $PAH$ emitting sources like Sh 2-61. Further details of these observations and associated errors are discussed by Ojha \etal (2018), Baug \etal (2018) and Sharma \etal (2020) along with science results. Baug \etal (2018) estimated 16.4, 14.0, 12.2 and 3.0 mag/arcsec$^{2}$ as sky brightness in $J, H, K$ and $nbL$ bands respectively. Similar values of sky brightness were also observed on October 7, 2017 during observations of the open star cluster Czernik (see Table 1). The values of sky brightness at Devasthal are comparable with those observed at other observatories like Hanle (Prabhu 2014), Calar Alto Observatory (Sanchez \etal 2008) and Las Campanas Observatory (Sullivan \& Simcoe 2012). At NIR wavelengths, Devasthal sky has been characterized for the first time. Because of its observational capability up to 3.59 \micron, the TIRCAM2 camera is extremely valuable instrument for observing those bright $nbL$ and $PAH$ sources which are saturated in the Spitzer-Infrared Array Camera observations.
 
 Anand \etal (2020) imaged two young planetary nebulae (PNe) namely NGC 7027 and BD +30\arcdeg 3639 in $J, H, K, PAH$ and $nbL$ filters of the TIRCAM2. These observations provided not only angular measurements of scientifically significant morphological features but also showed emissions from warm dust and $PAHs$ in the circumstellar shells of these young PNe. During these observations, the value of FWHM of a stellar image in $K$ pass band was derived as 0\farcsec{76} and 0\farcsec{62} on the nights of 13 October 2017 and 10 May 2018 respectively (see Table 1). 
 Using technique of lunar occultation for the first time on the 3.6 meter DOT, Richichi \etal (2020) carried out high angular resolution measurements of unresolved IRC-20156 and resolved SAO 98770 binary stars with the TIRCAM2. Based on these unique observations, Richichi \etal (2020) derived scientifically important astrophysical results. On both nights (18 and 21 May 2018), the value of FWHM of a stellar image in $K$ pass band was 0\farcsec{7} (see Table 1).  

 For the first time, on June 6, 2020, the 3.6 meter DOT successfully observed event of stellar occultation by a planet in $H$ pass band of the TIRCAM2. The star UCAC4 340-192403 was occulted by Pluto. Present observations are particularly important since occultations by Pluto to be seen from the Earth are getting rarer due to the fact that the dwarf planet now moves away from densely populated stellar regions of the Milky Way. Present high signal-to-noise ratio observations obtained as part of a global campaign are being analyzed and will be compared with models of Pluto’s atmosphere. A publication based on these detailed investigations is under preparation. 

\section{Summary and future outlook}

The values of FWHM estimated from images of stellar sources are listed in Table 1. They are expected to be slightly more than the value of atmospheric seeing prevailing at the epoch of observations as contribution from other sources of telescope image degradation discussed earlier in Section 2 are minimized by the use of the AOS. Atmospheric seeing is $\lambda$ dependent and varies as $\lambda^{-0.2}$ (refer Sagar \etal 2019b). One can therefore confidently state that at Devasthal, sub-arc-sec atmospheric seeing is routinely observed at visual wavelengths. The natural atmospheric seeing observed at the site $\sim$ 2 decades ago (see Sagar et al. 2000, and references therein), during 1997 to 1999, has, therefore, not deteriorated even after completion of the telescope and surrounding buildings. It is mainly because of the precautions taken in the design and the structure of the telescope building and locating the telescope floor $\sim$ 11 meter above the ground (Sagar \etal 2019a). All these coupled with use of low thermal mass material in the telescope building and installation of modern and complex ventilation facilities to keep minimum temperature gradient between telescope and its surroundings have paid a rich dividend by not deteriorated the natural atmospheric seeing at Devasthal. The 3.6 meter DOT can, thus, provide sky images with sub-arc-sec resolution at wavelengths ranging from optical to NIR for a good fraction of observing time. 
  
Performance verification of the telescope carried out during 2015-2016 after its installation at Devasthal indicates that all specifications of the 3.6 meter DOT given at the time of placing order in 2007 are successfully met. Its on sky imaging performance reveals that quality of its optics is excellent and capable of providing images of the celestial bodies with sub-arcsec (up to 0\farcsec4) resolutions. Observations taken with the IMAGER show that stars of $B = 24.5\pm0.2, R = 24.6\pm0.12$ and $ g =25.2\pm0.2$ mag have been detected in exposure time of 1200, 4320 and 3600 seconds respectively. In the photometric imaging mode of the ADFOSC, a distant galaxy of 24.3$\pm$0.2 mag and point sources of $25\pm0.3$ mag have been detected in the SDSS $i$ band in one hour of exposure time. The NIR observations taken with TIRCAM2 show that stars up to $J = 20\pm0.1,  H = 18.8\pm0.1$ and $K = 18.2\pm0.1$ mag can be detected in effective exposure times of 500, 550 and 1000 sec respectively. The TICAM2 is also capable of detecting the $nbL$ band sources brighter than $\sim$ 9.2 mag and strong ($\geq$ 0.4 Jy) $PAH$ emitting sources like Sh 2-61. At NIR wavelengths, Devasthal sky is now characterized with the TIRCAM2 observations. Both TANSPEC and TIRCAM2 at the focal plane of the 3.6-m DOT are very much suited for the search of low and very low mass stellar populations (M dwarfs, brown dwarfs), strong mass-losing stars on the asymptotic giant branch like young PNe (Anand \etal 2020) and young stellar objects still in their proto-stellar envelopes.

 The modern 3.6 meter DOT observing facilities  can provide optical and NIR observations for a number of front line Galactic and extra-galactic astrophysical research problems including optical follow up of $\gamma$-ray, X-ray, UV and radio sources observed with facilities like GMRT and AstroSat etc. and optical transients objects like SN and afterglows of $\gamma$-ray bursts and gravitational wave etc. These studies will ultimately contribute to our understanding of various types of terrestrial elements present in the periodic table created about 150 years ago in 1869 by  Russian chemist Dmitrii Mendeleev which revolutionized chemistry. 
 
 It is well known that the telescope only collects photons while the throughput of back end instruments built using latest cutting edge technology defines the quality of scientific output coming from it. It is essential to build modern and complex back end focal plane instruments so that full astronomical potential of the 3.6 meter DOT can be utilized. One such instrument named Devasthal Optical Telescope Integral Field Spectrograph (DOTIFS) is being fabricated by the Inter University Center for Astronomy and Astrophysics, Pune. The DOTIFS is a multi-object integral field spectrograph. The ARIES has recently initiated process of building a modern high resolution ($\sim$ 60,000) spectrograph equipped with most sensitive CCD detectors to be mounted at the telescope (Sagar \etal 2019b). Addition of these back end instruments will significantly increase scientific out put from the 3.6 meter DOT.
 
%%Use section* for acknowledgements
\section*{Acknowledgements}
 This article is based on the invited talk delivered during the 150 years of Periodic Table conference organized by the Indian Institute of Astrophysics (IIA), Benagaluru from 16 to 19 December 2019. Constructive comments provided by anonymous reviewer are very much appreciated. We are thankful to the staff of ARIES for their assistance during the observations with the telescope. One of us (Ram Sagar) thanks the National Academy of Sciences, India (NASI), Prayagraj for the award of a NASI honorary Scientist position; the Alexander von Humboldt Foundation, Germany for the award of Group linkage long-term research program and the Director, IIA for hosting and providing infrastructural support during this work. 
 
%%References section

\end{document}